\documentclass[reprint,superscriptaddress,twocolumn,aps,prl,showpacs,floatfix]{revtex4-1}
\usepackage{graphicx}
\usepackage{amsbsy,amssymb,amsmath,bm,ulem}

\begin{document}

\title{Ray optics in  flux avalanche propagation in superconducting films}

\author{P.~ Mikheenko}
\affiliation{Department of Physics,
University of Oslo, P. O. Box 1048 Blindern, 0316 Oslo, Norway}

\author{T. H. Johansen}
\affiliation{Department of Physics, University
of Oslo, P. O. Box 1048 Blindern, 0316 Oslo, Norway}

\affiliation{Institute for Superconducting and Electronic Materials,
University of Wollongong, Northfields Avenue, Wollongong, NSW 2522,
Australia}

\author{S. Chaudhuri}
\affiliation{Nanoscience Center, Department of Physics, P.O. Box
35, University of Jyv\"{a}skyl\"{a}, FIN-40014
Jyv\"{a}skyl\"{a}, Finland}

\author{I. J. Maasilta}
\affiliation{Nanoscience Center, Department of Physics, P.O. Box
35, University of Jyv\"{a}skyl\"{a}, FIN-40014
Jyv\"{a}skyl\"{a}, Finland}

\author{Y. M. Galperin}
\affiliation{Department of Physics, University
of Oslo, P. O. Box 1048 Blindern, 0316 Oslo, Norway}
\affiliation{Physico-Technical Institute RAS, 194021 St.~Petersburg,
Russian Federation}

\begin{abstract}

Experimental evidence of wave properties of dendritic flux
avalanches in superconducting  films is reported.
Using magneto-optical imaging the propagation of dendrites across boundaries 
between a bare NbN film and areas coated by a Cu-layer was visualized, 
and it was found that the propagation is refracted in full quantitative agreement 
with Snell's law. For the studied film of 170~nm thickness and a $0.9~\mu$m thick 
metal layer, the refractive index was close to $n=1.4$. The origin of the refraction
is believed to be caused by the dendrites propagating as an electromagnetic shock 
wave,  similar to damped modes considered previously for normal metals. The analogy 
 is justified by the large dissipation during the avalanches raising the local temperature 
significantly.  Additional time-resolved measurements of voltage pulses generated by 
segments of the dendrites traversing an electrode confirm the consistency of the 
adapted physical picture.

\end{abstract}


\maketitle

In thin-film superconductors placed in a gradually increasing or decreasing
transverse magnetic  field, the smooth propagation of the flux front
can be violated by sudden bursts of flux penetration.  These
dramatic events occur due to a thermomagnetic 
instability~\cite{rakhmanov04,aranson05,denisov05,denisov06} causing
large amounts of flux to rush in from seemingly arbitrary positions
along the edge. 
Magneto-optical imaging (MOI) of the flux penetration in  films of many superconducting
materials~\cite{leiderer93,duran95,johansen02,rudnev03,rudnev05,choi05,baziljevich14-1}
have shown that it is a generic feature of these avalanches that they
form non-repetitive branching structures,
see~\cite{altshuler04} for a review. It is also found
experimentally that the propagation speed
of such avalanches can be amazingly high, 
up to 160~km/s in the very early stage~\cite{bolz03,bolz03-2}.

In recent works~\cite{vestgarden12-sr, vestgarden14-sust}, 
new insight into the origin of such high
velocities was obtained by numerical simulations based on a set of coupled
equations for the electrodynamics and  thermal behavior of a superconducting 
film on a substrate.
The results show that due to the very large electrical fields and dissipation 
caused by rapid flux motion,
the local temperature  during such an avalanche typically rises above 
the superconducting  transition temperature.
One may therefore expect that in many superconducting compounds a propagating 
flux dendrite will show physical similarities with
electromagnetic modes in normal metal films~\cite{brandt94-strip,*brandt94-disk}.

In the present study of NbN films
we have searched for wave-like behavior of propagating flux dendrites.
In particular, samples were designed so that invading  dendrites will 
cross boundaries between two different superconducting media, represented 
by the bare NbN film and areas of NbN coated with a metal layer. It is well known that
a metal layer causes inductive braking of the avalanche 
propagation~\cite{colauto10,vestgarden14-sust}, and hence 
such boundaries should display refraction of dendrites
provided they have a  traveling wave nature.
The  paper gives direct experimental evidence, based on MOI observations, 
that  the electromagnetic modes excited in the dendritic avalanches
in NbN cause systematic refraction at
boundaries between different media. Moreover,  the quantitative refraction 
follows Snell's law,
in striking resemblance with geometrical optics of light.

\begin{figure}[b]
\centerline{
\includegraphics[width=\columnwidth]{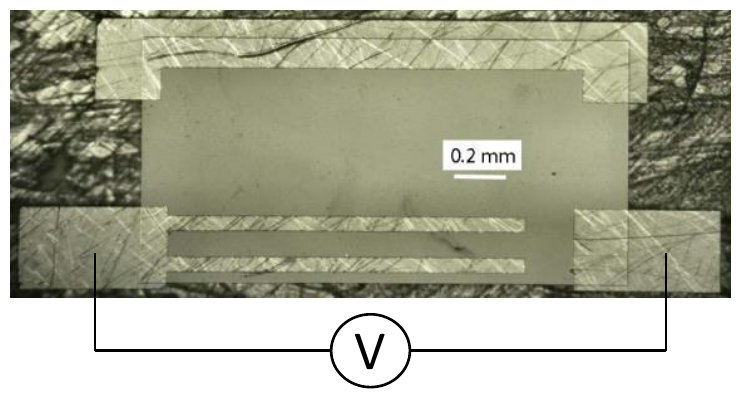}
} \caption{Optical image of the rectangular superconducting film equipped
with a pattern of metal coating. Shown is also the voltage pulse measurement 
circuit, which allows time-resolved observation of the avalanche dynamics. 
\label{sample}   }
\end{figure}

Films of NbN were grown on MgO(001) single
crystal substrate to a thickness of 170 nm
using pulsed laser deposition.
By electron beam lithography and reactive ion etching
with CF$_4$+O$_2$,  one film was shaped into a
3.0$\times$1.5~mm$^{2}$ rectangle.
Then, a 900~nm thick Cu layer was deposited on the film
and patterned as shown in Fig.~\ref{sample}.
Here, the two long horisontal strips of metal define areas where
flux avalanches starting from the lower edge will experience magnetic braking.
The metal coating along the upper edge of the film has the purpose of
preventing avalanches to start from that side of the sample.

In addition to MOI observations using a ferrite garnet indicator 
film~\cite{johansen96, helseth02},
the present work also makes use of the recent
finding~\cite{mikheenko13}  that a flux dendrite propagating in a 
metal-coated part of the superconductor generates a sizable voltage pulse.
To measure such pulses,  contact pads were placed at the lower corners of
the sample,  where the left pad contacts the two long metal strips.
With this geometry, if  two subsequent pulses are detected they provide
 information about the speed of the avalanche front.
Moreover, the fine structure of each pulse tells about the number of
flux branches passing the electrodes and the points in time they
enter and exit.
As voltmeter, a Tektronix TDS 210 oscilloscope was used, and
 set to record the voltage with sampling interval of a
few nanoseconds. The measurement was triggered by the pulse
 itself when the
instantaneous voltage exceeded a preset threshold value. When
triggered, the voltmeter stores also data measured in a
preceeding time interval, thus allowing the full profile of the voltage 
signal during an avalanche to be recorded.

\begin{figure}[t]
\includegraphics[width=\columnwidth]{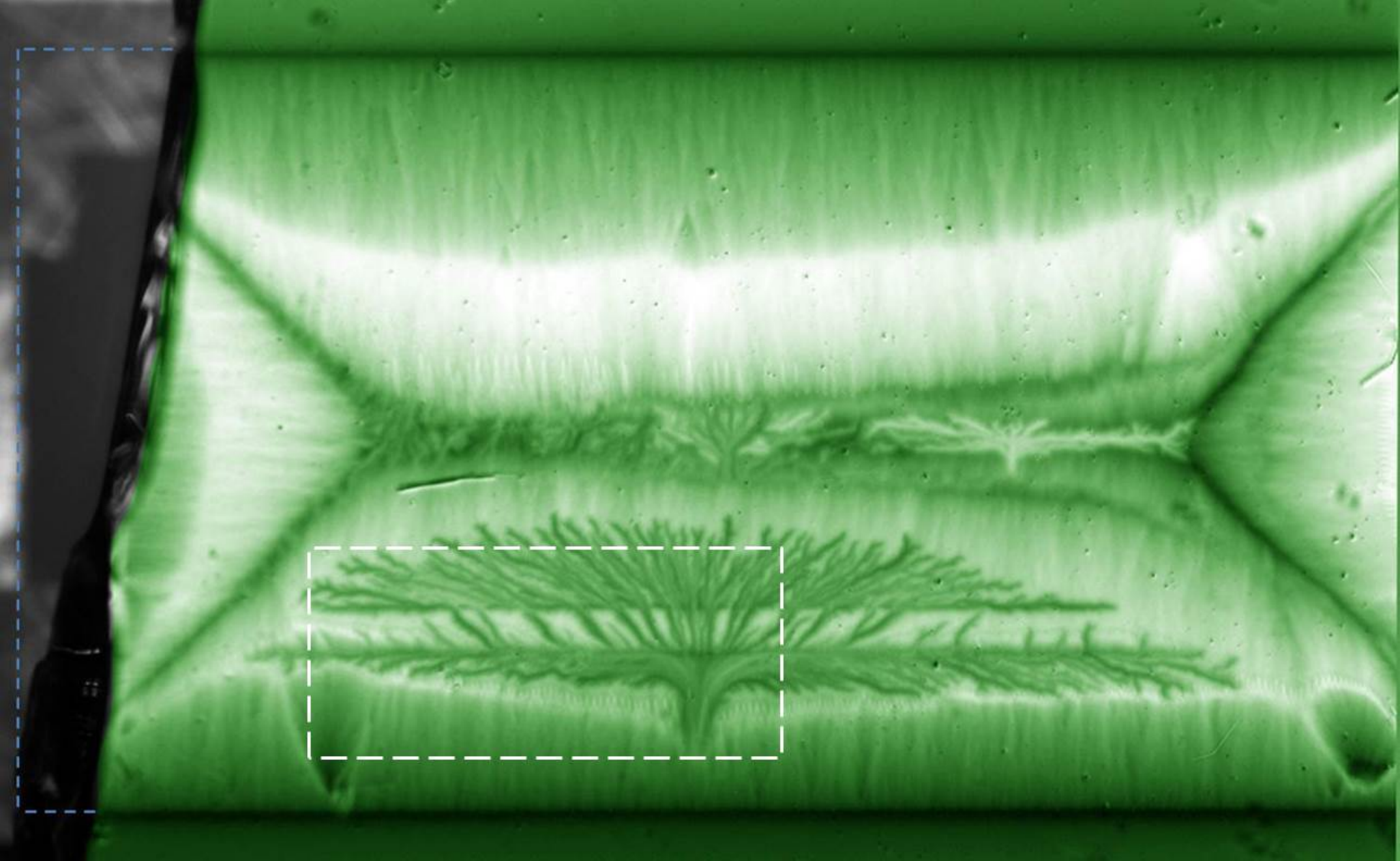}\\[2mm]
\includegraphics[width=\columnwidth]{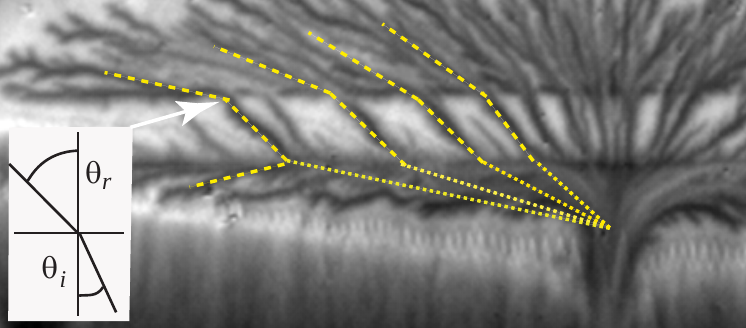}
 \caption{Magneto-optical image of a flux avalanche occurring at 3.7~K in the
metal coated NbN film.
The image covers the lower central part of the film, and was recorded
 in the remanent state after the field was first raised to 15~mT. 
The horizontal bright strip permeated by dark line 
segments is the metal coated strip located nearest the sample center.
The strip near the edge is invisible, as the avalanche crossed this region 
through a single channel perpendicular to the edge.
\label{fig2}}
\end{figure}

Shown in the upper panel of  Fig.~\ref{fig2}
is a magneto-optical image of 
the flux distribution after a typical  avalanche 
occurred in the NbN film at 3.7~K in descending applied magnetic field.
Prior to the field descent, the film was filled with flux by applying a 
perpendicular field of 17~mT, 
which removed essentially all the flux trapped from previous experiments,
and created an overall  flux distribution corresponding to a critical state. 
Then, during the subsequent field descent, when the field reached 14~mT,
 a large-scale  avalanche started from a location 
near the center of the lower sample edge.
The dark dendritic structure  shows the paths 
followed by antiflux as it abruptly invaded the sample. 
Note that in this transverse geometry the 
magnetic field around the film edge is oriented oppositely to the 
descending applied field, hence, the edge region becomes
 penetrated by antivortices during the descent.

Whereas the overall structure of this avalanche is quite complex, 
one can see that it consists of many branches, or rays, 
behaving with considerable regularity. 
Note from the image that one of the 
metal-coated strips  is seen here as the distinct bright region 
where only a few rays are crossing.
Note also that as long as the ray propagation takes place in the same medium, i.e., 
either the bare superconductor or the metal-coated superconductor, 
the rays are often quite straight. 
Moreover, when the rays traverse an interface between the two media,  
they show a clear refraction effect. 

To see this in more detail,  a magnified view of the 
flux distribution inside the rectangular area marked in Fig.~\ref{fig2}~(upper)
is shown in the lower panel. 
In the metal strip area the rays, indicated by dashed yellow lines,
 traverse the strip  at various angles denoted  $\theta_i$, 
see the insert for definitions. 
As the rays cross the interface
they continue into the bare superconductor at a different angle $\theta_r$.
This refraction angle is consistently larger than the
 incident angle, $\theta_i$. It is interesting to compare the two angles 
quantitatively in relation to  Snell's law,
$$ 
\sin \theta_r/ \sin \theta_i  = n, 
$$ 
where $n$ is the relative index of refraction of the metal-coated and bare 
areas of the superconductor. From the examples of refraction indicated by 
the dashed lines in Fig.~\ref{fig2}~(lower) one finds $n = 1.37, 1.37, 1.44$ 
and 1.34, which are remarkably similar values. 

These observations give strong indications that the avalanche dynamics is 
governed by oscillatory electromagnetic modes, 
and that these modes have different propagation velocities in the bare superconductor and 
metal-coated film. Denoting these two velocities $v_s$ and $v_c$, respectively, 
the suggested physical picture then demands that their ratio is equal to 
the index of refraction, $v_s/v_c =n$. This relation was tested
 by analyzing additional experimental data.

\begin{figure}[t]
\centerline{
\includegraphics[width=\columnwidth]{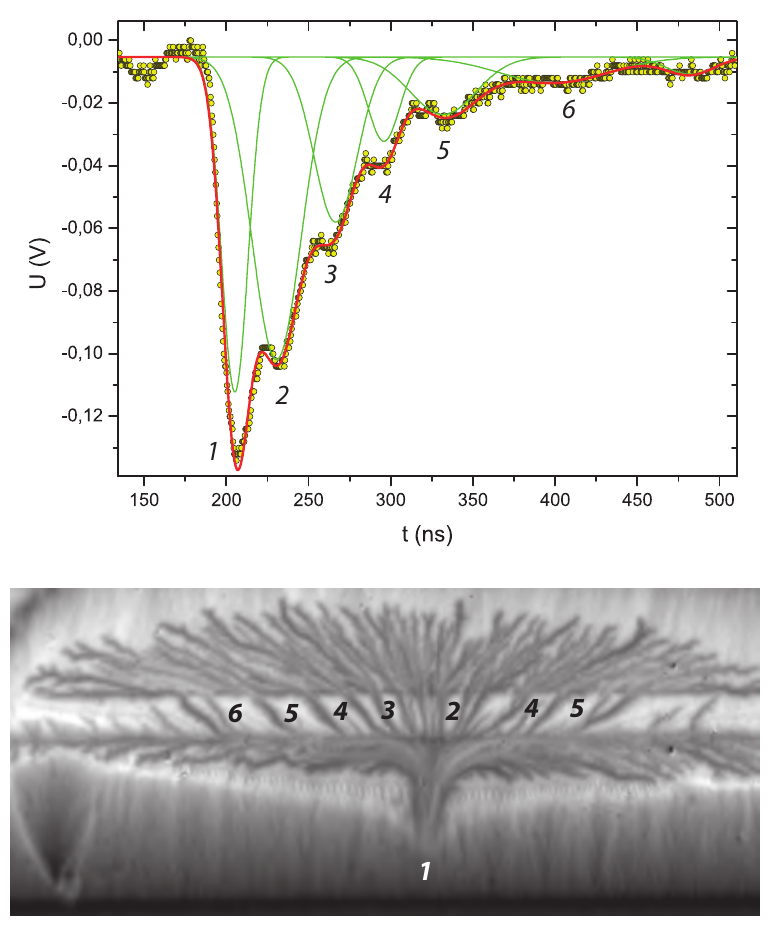}
} \caption{Voltage pulse generated by the flux avalanche 
seen in the magneto-optical image below. 
The signal is decomposed into a sequence of shorter pulses (green curves)
 numbered from 1 to 6, which give rise to corresponding 
peaks in the overall curve. See the text for the numbers assigned to the
peaks and dendrite segments in the image.\label{fig3}}
\end{figure}

During the avalanche event seen 
in Fig.~\ref{fig2}, also the voltage signal it generated between the 
electrodes was recorded, and
the result is shown in the upper panel of Fig.~\ref{fig3}. 
The avalanche created a pulse of 200~ns duration and 
maximum magnitude 0.14~V.
From the profile it is evident that the signal can be 
considered as composed of  a set of
shorter subsequent pulses. As shown in previous work~\cite{mikheenko13}, 
such a pulse is observed when a flux dendrite propagates across the electrode area. 
The decoding of the measured signal is therefore possible using the information
provided by the magneto-optical image in the lower panel.

The first peak in the voltage signal was generated when 
the lower metal strip 
(not clearly seen in the magneto-optical image)
was traversed by the full avalanche front. 
The first peak has a maximal value of 0.11~V, and a width close to 20~ns. 
All the subsequent peaks are caused by dendrites traversing the upper strip electrode.
It is here reasonable to assume that the dendrites nearest the main trunk, i.e., 
those marked in the lower panel by ``2", were the first to 
cross the upper electrode, and hence 
give rise to the peak marked ``2" in the voltage signal. 
Marked in both panels by increasing numbers is our 
suggested correspondence between subsequent peaks 
and dendrite segments crossing the upper electrode. 
This is not a detailed one-to-one correspondence since  
several dendrites may contribute to the same peak.

Let therefore the quantitative analysis take into account only the first two peaks. 
The time delay between them is $\Delta t =25$~ns.
 During this interval the avalanche propagates a distance
close to the sum of two strip electrode half-widths, each $w =0.045$~mm,
plus the separation between them, $s=0.16$~mm.
The delay time can therefore be expressed as
$$
\Delta t =   2\frac{w}{v_c} + \frac{s}{v_s} \ .
$$
Combining this with $v_s/v_c =n$ where the ray refraction observations gave as 
average, $n = 1.38$, 
we find that the velocity of dendrite propagation in the metal-coated part of the NbN film 
is $v_c = 8.2$ km/s. In the bare superconducting film the velocity is $v_s =11.5$~km/s.
This quantitative result for the dendrite propagation speed at an intermediate stage 
of an avalanche is fully consistent with previous measurements near the final stage 
of an avalanche in a similar film, which gave a value of 5~km/s~\cite{mikheenko13}.

The surprising observation that flux dendrites propagating across
 boundaries between two  superconducting media show quantitative 
agreement with Snell's refraction law 
leads us to conclude that the dendrites propagate electromagnetic modes 
with well-defined speed. Such modes propagating in a film of resistivity
$\rho$ were considered in~\cite{brandt94-strip,*brandt94-disk}, where it was found that
 their speed    can be written as,
\begin{equation} \label{vem}
v_{\text{em}}=\alpha \rho /\mu_0 d .
\end{equation}
Here, $d$ is the sample thickness, $\alpha \simeq 1$ is a 
numerical factor depending on the sample geometry and type of  mode,
and $\mu_0$ is the vacuum magnetic permeability. 

For a superconducting film of thickness $d_s$ and resistivity, $\rho_s$, 
coated by a metal layer of thickness $d_m$ and resitivity $\rho_m$,
one can define an effective resistivity $\rho_c$.  
If there is no exchange of electrical charge between the two layers, the 
resistivity of the coated film is given by,
\begin{equation}\label{effcond}
\rho_c =  (d_s+d_m) \, \left( \frac{d_s}{\rho_s}+ \frac{d_m}{\rho_m}\right)^{-1}\, .
\end{equation}
From Eq.~(1) it then follows that the propagation velocity in the bare superconducting film, 
$v_s$,  and the velocity  in the coated film, $v_c$, are related by
\begin{equation} \label{n1}
\frac{v_s}{v_c} = 1 + \frac{\rho_s \,  d_m}{\rho_m \,  d_s} \, .
\end{equation}

Thus, from Snell's law,  the relative refractive index for rays propagation between 
coated and bare areas of a superconducting film is given by the rhs of Eq.~(\ref{n1}).
The ratio,  $  (\rho_s  d_m)/(\rho_m \,  d_s)   \equiv S  $,
was introduced recently~\cite{vestgarden14-sust} as a parameter 
to quantify how efficiently a metal coating  will suppress flux avalanches in an adjacent superconductor.  
Using again $n =1.38$, we find for the present system that $S=0.38$.
Compared with the case considered in \cite{vestgarden14-sust}, where 
$S\gg 1$ and the metal coating caused rapid decay of the avalanches,
the present $S$-value represents weak damping, which evidently is a prerequisite for 
refraction of the dendrites to be observed.

With the values for $d_s$ and $d_m$ in the present sample, 
one finds $\rho_s \approx 0.07 \rho_m$. 
From this it follows that the instantaneous temperature 
at the front of a propagating dendrite is not far from the 
superconductor's critical temperature. 
Also this is  consistent with the assumption that the front 
 propagation can be considered 
analogous to the modes introduced  in~\cite{brandt94-strip,*brandt94-disk}. 

To visualize the refraction in the bare superconductor below the upper strip,
we show in the lower panel of Fig.~\ref{fig2} drawing of straight dotted lines
parallel to the refracted rays in the bare region above the strip. 
Interestingly, the dotted lines cross each other at one single point. 
This particular location appears to be the spot from where
the massive initial avalanche splits into many branches. 
However, distinct branches are not clearly resolve in this region. 
We interpret this to be a result of the large
initial heating in this "root" region of the avalanche making the 
branches blurred.
In the same panel one can make another intersting observation,
namely a clearly visible example of  dendrite reflection. The event takes place
at the lower edge of the strip, and the reflected ray is drawn as a dashed line 
at an angle equal to that of the incident ray.

From the present work, we conclude that
the propagation of thermomagnetic avalanches in the form of distinct branches
observed by MOI has significant similarities with
 the propagation of an optical ray through interfaces between two media. 
 By quantitative comparisons 
it was demonstrated that the branches undergo refraction in full agreement with Snell'slaw. 

These findings support an interpretation of the flux front dynamics at its fast initial stage as 
 propagation of a damped shock electromagnetic wave of the kink type, see, 
e.g.,~\cite{Shock_waves}.  We ascribe these waves to damped electromagnetic 
modes similar to those considered in~\cite{brandt94-strip,*brandt94-disk}.
Usually, shock waves contain many Fourier components.  If the medium is 
 dispersive, i.e., the velocities of different component are essentially different, 
the refracted ray gets smeared.  Within the present resolution, this is not 
observed here, and we conclude that 
the velocities of the different Fourier components form a narrow distribution.

To observe ray-like refraction of flux dendrites one needs superconducting films where the
propagation velocities in the different parts of the device is not too different.  
For a partly metal-coated film the braking parameter 
$S$ should be not too large, otherwise the damping will dominate. We leave it for future 
work to identify more precisely the boundaries in parameter space
of the interesting low-damping regime reported in this paper.

\acknowledgments
The financial support of the Research Council of Norway was gratefully acknowledged. Research at the University of  Jyv\"{a}skyl\"{a} was supported by the Academy of Finland within Project No. 128532.

%

\end{document}